\documentclass[aps,pra,twocolumn,
epsfig,showpacs,superscriptaddress,footnote,nofootinbib]{revtex4-1}
\usepackage{amsmath,amsthm,amssymb,mathrsfs,amsfonts,dsfont}
\usepackage{graphicx}
\usepackage{color}
\usepackage{multirow}
\usepackage{wrapfig}
\makeatletter
\def\be{\begin{equation}}
\def\ee{\end{equation}}
\def\bea{\begin{eqnarray}}
\def\eea{\end{eqnarray}}
\def\ben{\begin{equation*}}
\def\een{\end{equation*}}
\def\bean{\begin{eqnarray*}}
\def\eean{\end{eqnarray*}}
\def\bma{\begin{mathletters}}
\def\ema{\end{mathletters}}
\def\bi{\begin{itemize}}
\def\ei{\end{itemize}}
\def\bd{\begin{description}}
\def\ed{\end{description}}

\newcommand{\ket}[1]{| \, #1 \rangle}

\begin{document}
\title{GHZ correlation provides secure Anonymous Veto Protocol}
\author{Ramij Rahaman}
\email{ramijrahaman@gmail.com} \affiliation{Department of Mathematics, University of Allahabad, Allahabad 211002, U.P., India}
\author{Guruprasad Kar}
\email{gkar@isical.ac.in} \affiliation{Physics and Applied Mathematics Unit, Indian Statistical Institute, 203 B T Road, Kolkata 700108, India}


\begin{abstract}
Anonymous Veto (AV) and Dining cryptographers (DC) are two basic primitives for the cryptographic problems where the main aim is to hide the identity of the senders of the messages. These can be achieved by classical methods where the security is based either on computational hardness or on shared private keys. In this regard, we present a secure quantum protocol for both DC and AV by exploiting the GHZ correlations. We first solve a generalized version of the DC problem with the help of multiparty GHZ state. This allow us to provide a secure quantum protocol for the AV. Securities for both the protocols rely on some novel and fundamental features of GHZ correlations related to quantum nonlocality.
\end{abstract}

\maketitle
\section{Introduction} In the classical world, where any physical transmission can be traced to its origin, it seems impossible to setup a secure way for message transmission without revealing its senders' identity. Dining cryptographers (DC) problem \cite{Chu88} introduced by Chaum is one of the primary attempts in this context. In a DC problem, three cryptographers are curious to find out whether their agency NSA (U.S. National Security Agency) or one of them pays for the dinner. At the same time they respect each other's right to make an anonymous payment. A generalized version of the DC problem called DC-net where one of the member from an agency publicizes a secret message without revealing his identity \cite{Chu88}. An unconditionally secure DC-net requires pairwise shared (secure) keys and an authenticated broadcast channel. Since, the security of DC-net relies on the generation of secure key between pairs of members so it is not {\em unconditionally secure}\footnote{security based on {\em computational hardness} of same nature as classical key distribution protocol.} if members are not allowed to pre-share bilateral private keys. Another major flaw of DC problem is that, it is vulnerable against multiple payments. It shows zero pay {\em i.e.} no transmission of message if even number ($0,2,\ldots$) of members pay for the dinner and detects payment if an odd ($1,3,\ldots$) number of members pay for the dinner. There is another loophole in DC problems called collusion loophole where some of the participants may cooperate among them to trace the person who pays. There are some works that partially resolve the problem with multiple payments and the collusion problem but none of them provides an unconditionally secured solution \cite{KY03,Gro04,Bra05}.

Another variant of DC problem known as {\em Anonymous Veto (AV)} problem \cite{HZ06}. Here a group of jury members, who need to take an unanimous decision, but at the same time want their individual decisions to remain secret {\em i.e.} without ever disclosing the identity of possible vetoing member(s). This could be very important in many aspects of human societies. Security of the classical solution of this problem is also based either on the {\em computational hardness} like other classical cryptographic protocols or on imposing restrictions on the number of dishonest players \cite{BT07}. In this context, Boykin \cite{Boy02} provided a quantum protocol to send classical information anonymously by distributing pairwise shared {\em EPR pairs}\footnote{two-qubit maximally entangled state $\frac{1}{\sqrt{2}}\left[\ket{00}-\ket{11}\right]$} among players. In 2005, Christandl and Wehner \cite{CW05} proved that the protocol presented by Boykin is not perfectly secure since it does not satisfy the traceless property and they provided an alternative quantum scheme of the DC-type problem with the traceless feature. In this regard, we present secure quantum protocol for both DC and AV problems with the help of multi-qubit GHZ correlation and GHZ paradox \cite{GHZ89}.

We start with a brief description of the GHZ paradox, which will allow us to present a quantum protocol for the three-party DC problem with a detection of multiple payments. This three party DC protocol is quite similar to the protocol presented in \cite{CW05}. We then extend the protocol into $n$-party DC problem without any detection of multiple payments. By exploiting this generalized version of DC problem we demonstrate a quantum protocol for the AV problem.
\section{GHZ paradox} In 1989, Greenberger, Horne and Zeilinger (GHZ) \cite{GHZ89} provided a way to show a direct contradiction
of quantum mechanics with local realism without using any statistical inequality. Consider a three qubit {\it maximally entangled}\footnote{this entanglement is maximal in the sense that
it gives the maximum violation of Bell's inequality for a given set of observables} \be\label{GHZ}\ket{\Psi}= \frac{\ket{000}-\ket{111}}{\sqrt{2}}\ee known as GHZ states. This GHZ state satisfies the following four constraints
\bea\begin{split}\label{GHZP}
\sigma_x\otimes\sigma_x\otimes\sigma_x\ket{\Psi}&=(-1)\ket{\Psi}\\ \sigma_x\otimes\sigma_y\otimes\sigma_y\ket{\Psi}&=(+1)\ket{\Psi}\\
\sigma_y\otimes\sigma_x\otimes\sigma_y\ket{\Psi}&=(+1)\ket{\Psi}\\ \sigma_y\otimes\sigma_y\otimes\sigma_x\ket{\Psi}&=(+1)\ket{\Psi},
\end{split}\eea
where, $\sigma_x,\sigma_y,\sigma_z$ are the Pauli matrices. Then, one can easily show that the above four constraints cannot be satisfied simultaneously by any local-realistic (LR) theory \cite{GHZ89,Mer90}. Similarly, for another three-qubit GHZ state \be\label{GHZ1}\ket{\Psi^{\perp}}= \frac{\ket{000}+\ket{111}}{\sqrt{2}}\ee we have,
\bea\begin{split}\label{GHZP1}
\sigma_x\otimes\sigma_x\otimes\sigma_x\ket{\Psi^{\perp}}&=(+1)\ket{\Psi^{\perp}}\\
\sigma_x\otimes\sigma_y\otimes\sigma_y\ket{\Psi^{\perp}}&=(-1)\ket{\Psi^{\perp}}\\
\sigma_y\otimes\sigma_x\otimes\sigma_y\ket{\Psi^{\perp}}&=(-1)\ket{\Psi^{\perp}}\\
\sigma_y\otimes\sigma_y\otimes\sigma_x\ket{\Psi^{\perp}}&=(-1)\ket{\Psi^{\perp}}.
\end{split}\eea
Like the previous case, the above four constraints also cannot be satisfied simultaneously by any LR theory.
\section{Quantum Dining Cryptographers (QDC) protocol} Imagine that three cryptographers Alice, Bob and Charlie want to play the Dining Cryptographers (DC) problem. To do that they first share a number (say $L_1$) of copies of the GHZ state $\ket{\Psi}$ given in (\ref{GHZ}), one qubit each from each copy. Here a copy of the states corresponds to a run of the protocol. Onward we use both the notations {\em copy of the state} or {\em run of the protocol} synonymously. After receiving all the qubits from $L_1$ copies of GHZ states they randomly select some runs (say $L_2$) and check whether the selected states satisfy the GHZ paradox or not. If yes, rest of the shared states (say, $L=L_1\smallsetminus L_2$) are genuine copies of GHZ state (\ref{GHZ}). The detail of the genuineness check of GHZ state is discussed discussed later. After confirmation of genuineness of the states the protocol goes as follows:\\
\begin{tabular}{|l|}
 \hline
 Protocol: QDC(3) \\
 \hline
 S1. Each member performs $\sigma_z$ on his qubits if he wants\\~~~ to  pay for the dinner otherwise does nothing. \\
 S2. Randomly select a copy of the states to distinguish\\~~~ between the cases (i) even and (ii) odd no. of payment(s).\\
 S3. Distinguish `no pay' vs. `double pay' in case (i)\\~~~ and  `single pay' vs. `triple pay' in case (ii).\\
\hline
\end{tabular}
\bd
\item[S1] {\em Performing local unitary operation to encode payment:} Alice performs local unitary operation $\sigma_z$ on each of her qubits from $L$ if she wishes to pay the dinner. Otherwise, she does nothing. Bob and Charlie follow the same.
\begin{description}
\item[(i)] If an even number ({\em i.e.} zero/two) of members pay the bill ({\em i.e.}, apply $\sigma_z$) then the states of all the members of $L_2$ remain in the same GHZ state (\ref{GHZ}) as
\ben\begin{split}
\mathds{I}\otimes\mathds{I}\otimes\mathds{I}\ket{\Psi}=\ket{\Psi};&~ \mathds{I}\otimes\sigma_z\otimes\sigma_z\ket{\Psi}=\ket{\Psi};\\
\sigma_z\otimes\mathds{I}\otimes\sigma_z\ket{\Psi}=\ket{\Psi};&~ \sigma_z\otimes\sigma_z\otimes\mathds{I}\ket{\Psi}=\ket{\Psi}.
\end{split}\een
\item[(ii)] If an odd number ({\em i.e.} one/three) of members want to pay for the dinner, the states are transformed to $\ket{\Psi^{\perp}}$ given in (\ref{GHZ1}) as
\ben\begin{split}
\sigma_z\otimes\mathds{I}\otimes\mathds{I}\ket{\Psi}=\ket{\Psi^{\perp}};&~\mathds{I}\otimes\sigma_z\otimes\mathds{I}\ket{\Psi}=\ket{\Psi^{\perp}};\\
\mathds{I}\otimes\mathds{I}\otimes\sigma_z\ket{\Psi}=\ket{\Psi^{\perp}};&~\sigma_z\otimes\sigma_z\otimes\sigma_z\ket{\Psi}=\ket{\Psi^{\perp}}.
\end{split}\een
\end{description}
In the next step of the protocol members distinguish between two cases (i) and (ii) without disclosing payer(s) identity.
\item[S2] {\em Distinguishing case (i) and case (ii):} To distinguish case (i) and case (ii), members randomly select one of the run (say $r_1$-th) from $L$. Now the task is to identify the state ($\ket{\Psi}$ or $\ket{\Psi^{\perp}}$) corresponding to the run $r_1$. Thus, the distinguishability task between case (i) and case (ii) reduces to the problem of distinguishability of two orthogonal three-qubit GHZ states $\ket{\Psi}$ and $\ket{\Psi^{\perp}}$ \cite{WH02}.

Now the task is to distinguish between subcases `{\em zero pay}' vs. `{\em double pay}' for case (i) and `{\em single pay}' vs. `{\em triple pay}' for case (ii).
\item[S3] {\em Distinguishing between subcases:} Members randomly selects one of the run (say $r_2$-th) from $L\smallsetminus \{r_1\}$. If case (i) occurs in the previous step then each member measures $\sigma_y$ on his qubit if (s)he pays for the dinner otherwise measures $\sigma_x$. If the product of the local measurements is $-1$ then no member has paid for the dinner ({\em i.e., zero pay}) and if the product is $+1$ then two members have paid for the dinner ({\em i.e., double pay}). The first case follows from the top equation of (\ref{GHZP}) whereas, the second case follows from the last three equations of (\ref{GHZP}). If {\em zero pay} occurs NSA will pay for the dinner and if {\em double pay} occurs the payment will be cancelled.

 In case (ii), each member measures  $\sigma_x$ on his qubit if (s)he pays for the dinner otherwise, measures $\sigma_y$. If the product of the local measurements is $+1$ then all the three members have paid for the dinner ({\em i.e., triple pay}) and if the product is $-1$ then only one member has paid for the dinned ({\em i.e., single pay}). The first case follows from the first equation of (\ref{GHZP1}) whereas, the second case follows from the last three equations of (\ref{GHZP1}). If a {\em single pay} occurs payment will be accepted otherwise payment will be cancelled.
\ed
Obviously, two copy of the states, one copy for each step, are sufficient. Therefore, it is enough if the list $L$ contains just two runs. This is true only if all the members honestly follow the entire protocol {\em i.e.,} perform local measurements (consistently) whenever asked according to their action and declare the true outcomes for each such measurement. If they act dishonestly, they do it solely to trace payer(s) identity only, and not to create any confusion regarding payment. The member who announces his results last in both the steps S2 and S3 enjoys some advantage. (S)He may change the case by just sending a flipped result of her/his measurement outcome. To deal with this problem, members choose more than one run in both steps S2 and S3 and the ordering of the announcement of the result is made random for each such selected run in both steps. Therefore, no members gets the advantage of being last to announce and if someone still flips the result, that will lead to an inconsistent conclusion and subsequently, they abort the protocol and starts a new one with a fresh set of resources.
\subsection{Security analysis of QDC protocol} Since the payer(s) information in step S1 encoded inside the phase of the GHZ state and due to the party symmetry of the state no quantum operation can reveal the identity of the payer(s). So the local operations for distinguishing the cases never disclose any information about the payer(s) identity. Step S2 only disclose the information whether the total number of payers are odd or even. This information in no way harm the purpose, rather it helps to detect multiple payments. In step S3 members only reveal their individual measurement result and not the choice of measurement to identify the `{\em no pay}' in case of (i) and a `{\em single pay}' in case of (ii). By knowing measurement result one cannot predict the measurement choice as that would immediately imply a violation of causality principle. If two of the members cooperate to each other then they can certainly predict the measurement choice and hence the action of the third party by knowing the measurement result. But this is quite obvious, since the anonymity exists only among a set of possible performers, and if the set is singleton, its member is always traceable {\em i.e.,} no protocol can keep the singleton members set untraceable. In our QDC protocol if the payment accepted {\em i.e.} a `{\em single pay}' happens then no non-payer have any information about the payer. But, in case of rejection of payment the identity of payers may be disclosed in two cases (i) if the `{\em double payment}' occurs then the non-payer knows that the other two members are the payers. This can be avoided if we assume that payment is made by one member only like the original DC problem and then the protocol will end at step S2. Based on the assumption that multiple payments will never occur, one can easily generalize our QDC protocol for $n(\geq 3)$ number of members.
\subsection{Generalized QDC protocol} Let a group of $n$ cryptographers are sitting for dinner at a restaurant and they want to find out whether their agency NSA or one of them pays for the dinner, while respecting each other's right to make a payment anonymously. To implement the protocol $n$-cryptographers share a copy of the generalized n-qubit GHZ state
\be\label{GHZg}
\ket{\Psi_n}=\frac{1}{\sqrt{2}}\left[\ket{000\ldots 0}-\ket{111\ldots 1}\right].
\ee
The above GHZ state has the following correlation:
\bea\label{GHZg1}
\sigma_z^{\bar{t}}\ket{\Psi_n}&=& \frac{1}{\sqrt{2}}\left[\ket{000\ldots 0}-(-1)^t\ket{111\ldots 1}\right]\nonumber\\
&=&\frac{1}{\sqrt{2}}\left[\ket{000\ldots 0}-\ket{111\ldots 1}\right]=\ket{\Psi_n}\nonumber\\
&&~~~~~~~~~~~~~~~~~~~~~~~~~~~~~~~~~~ \mbox{ (if $t$ is even)},\nonumber\\
&=&\frac{1}{\sqrt{2}}\left[\ket{000\ldots 0}+\ket{111\ldots 1}\right]=\ket{\Psi_n^{\perp}}\nonumber\\
&&~~~~~~~~~~~~~~~~~~~~~~~~~~~~~~~~~~\mbox{ (if $t$ is odd)},
\eea
where $\sigma_z^{\bar{t}}$ denotes that in $t$-number of places $\sigma_z$ acts and in rest of the places $2\times 2$ identity matrices $\mathds{I}$ acts. For the generalized DC problem with no multiple payments $t=0,1$.
Same kind of protocol with step S1 and S2 as described above will work in this case. By exploiting this generalized QDC protocol we now provide a secure quantum protocol for the AV problem for odd number of parties. Then we extend the result for even number of parties.
\section{Quantum Anonymous Veto (QAV) protocol} Imagine a jury with $n$ members, who need to take an unanimous decision, but at the same time want their individual decisions to remain secret. The generalized GHZ state $\ket{\Psi_n}$ given in (\ref{GHZg}) would allow them to achieve this. The quantum AV protocol starts with sharing $L$ ($L\geq 2$) genuine\footnote{To check the genuineness of states they randomly select some copies of them and run the GHZ-type paradox as described in section \ref{gen}.} copies of $\ket{\Psi_n}$ between jury members. Each member gets one qubit from each of the copy of $\ket{\Psi_n}$.

\begin{tabular}{|l|}
 \hline
 Protocol: QAV (odd-$n$) \\
 \hline
 S1'. Each member performs $\sigma_z$ on his qubits if he\\~~~~~~ wants to vote `against'
 otherwise, does\\~~~~~~  nothing. \\
 S2'. Performs $\sigma_x$ on qubit associated to a selected\\~~~~~~  run to distinguish between
 the cases \\~~~~~~ (i) even (including zero) and (ii) odd no. of \\~~~~~~ `against' votes.\\
 S3'. For case (i), distinguish between the cases of\\~~~~~(a) unanimity `in favor' and (b) an even\\~~~~~~(excluding zero) no. of `against' votes.\\
 \hline
\end{tabular}
\bd
\item[S1'] After receiving all the qubits, each member performs the unitary operation $\sigma_z$ if he is `against' the decision and does nothing if he is `in favor'. (i) If an even number of members (including zero) vote `against' the decision, all the states remain same as $\ket{\Psi_n}$. (ii) Otherwise, all the states transform to $\ket{\Psi_n^{\perp}}$.
\item[S2'] Jury members randomly select one copy of the state to distinguish between the cases (i) and (ii) by distinguishing two orthogonal states \cite{WH02}. Unanimity in favor of the decision happens only if no members ({\em i.e.} zero members) voted against. Since, (ii) represents the case where at least one of the members voted against so it does not require any farther analysis. But, case (i) represents (a) the unanimity `in favor' of the decision and (b) an even number ($2,4,\ldots$) of members voted against the decision.
\item[S3'] To distinguish between subcases (a) and (b) they first randomly select one copy of the state and each member performs (again) the unitary operation $\sigma_z(1)=\left(
 \begin{array}{cc}
1 & 0 \\
0 & \cos\frac{\pi}{2}+i\sin\frac{\pi}{2} \\
\end{array}
\right)
$ on his qubit if he is `against' the decision, otherwise does nothing. (i) If the number (including zero) of members against the decision is even multiple of $2$ ({\em i.e.}, multiple of $2^2$) then the selected state will remain in $\ket{\Psi_n}$. (ii) Otherwise, ({\em i.e.}, the number of members against the decision is odd multiple of $2$) it will transform to $\ket{\Psi_n^{\perp}}$.

After distinguishing between these two cases, further analysis has to be made for case (i) in S3.

Case (i) represents (1a) the unanimity in favor of the decision and (1b) multiple of $2^2$ ({\em i.e.}, $4,8,12,16,20,\ldots$) no. of members voted against the decision.

To distinguish between subcases (1a) and (1b), they again select another copy and each party perform the following unitary operation $\sigma_z(2)=\left(
 \begin{array}{cc}
1 & 0 \\
0 & \cos\frac{\pi}{2^2}+i\sin\frac{\pi}{2^2} \\
\end{array}
\right)
$ on his respective qubit if he is `against' the decision, otherwise does nothing. If an even multiple of $2^2$ no. ({\em i.e.}, multiple of $2^3$) of members (including zero) are against the decision the copy remains unchanged. Otherwise, ({\em i.e.}, an odd multiple of $2^2$ no. of members excluding zero are against the decision) the copy transforms to $\ket{\Psi_n^{\perp}}$. Again these two cases can be distinguished by distinguishing the two orthogonal states.

Jury members keep repeating these steps. In general, to distinguish the case of even multiple of $2^t$ no. of members (including zero) favouring the decision and the odd multiple of $2^{t}$ no. of members against the decision, the required unitary operation will be $\sigma_z(t)=\left(
 \begin{array}{cc}
1 & 0 \\
0 & \cos\frac{\pi}{2^t}+i\sin\frac{\pi}{2^t} \\
\end{array}
\right)
$.
Since, the total number of jury members are finite so after a finite number of steps they can detect whether there is any unanimity `in favor' of the decision.

Note that in the entire protocol the identity of the member giving veto(es) is not revealed. The thing that is revealed is the information regarding the number of vetoes. Here also the security is guaranteed from the genuineness of the GHZ states.
\ed
If the number of jury members $n (>2)$ is even then members share copies of $\ket{\Psi_{n+1}}$ where one (say, first) of the jury members holds two qubits from each copy. Here, except the first jury member all the other members follow the similar protocol as described in case of odd no. of members. In each run, the first member treats the first qubit (from the pair of qubits he received at each run) as earlier {\em i.e.,} performs operation/measurement according to his choice of decision and on the second qubit he always performs the operations according to decision `in favor'. Obviously, this arrangement does not provide any advantage to the first member and hence does not effect the objectivity of the protocol.
\section{Genuineness check of GHZ state}\label{gen} Security of all the protocols described above is solemnly dependent on the genuineness of the corresponding GHZ state. Since, one can construct a secure even $(n-1)$-parties QAV(QDC) protocol from an odd $n$-parties QAV(QDC) protocol so here we describe only the genuineness check of GHZ states for odd $n$. To check the genuineness of n(odd)-qubit GHZ state each player $j (j=1,2,\cdots,n)$ randomly selects some runs $R_j$ (i.e. copies of the shared n-qubit GHZ state) and for each $r\in R_j$ he again randomly chooses a operator $\mathcal{O}_{j_r}\in\{\mathcal{O}_i\}_{i=0}^n$, where
\ben
\mathcal{O}_0=\sigma^1_x\otimes\cdots\otimes\sigma^{i-1}_x\otimes\sigma^i_x\otimes\sigma^{i+1}_x\otimes\sigma^{i+1}_x\otimes\cdots \otimes\sigma^n_x\mbox{  and}
\een
\ben
\mathcal{O}_i=\sigma^1_x\otimes\cdots\otimes\sigma^{i-1}_x\otimes\sigma^i_y\otimes\sigma^{i+1}_y\otimes\sigma^{i+1}_x\otimes\cdots \otimes\sigma^n_x,
\een for $i=1,2,\cdots,n$ with the convention $n+1\equiv 1$. The upper indices on Pauli matrices represent the identity of the party. Now player $j$ asks player $t (t=1,2,\cdots,n)$ to measure his qubit (associated with the run $r$) in the basis that present in the $t$-th place of the operator $\mathcal{O}_{j_r}$ and send the measurement outcome. Player $j$ collects all the local measurement data (including his won measurement result) corresponding to the operator $\mathcal{O}_{j_r}$ and checks whether the product of the local measurement results is equal to the eigenvalue $\lambda_{j_r}$ of the eigenvalue equation
\bea\begin{split}\label{GHZPn}
\mathcal{O}_{j_r}\ket{\Psi_n}&=\lambda_{j_r}\ket{\Psi_n},
\end{split}\eea
where $\lambda_{j_r}=-1$ if $\mathcal{O}_{j_r}=\mathcal{O}_{0}$ otherwise, $\lambda_{j_r}=+1$.
The above relations provide a GHZ like contradiction with LR-theory for an n-qubit system when $n$ is odd. By employing relations given in (\ref{GHZPn}), one can construct the following LR inequality for $n$-(odd) two level system.
\be\label{GHZPnI}
\mathcal{O}=\left|\left\langle\sum_{i=1}^n\mathcal{O}_i-\mathcal{O}_0\right\rangle\right|\leq (n-1).
\ee
The two extreme eigenvalues of the operator $\left(\sum_{i=1}^n\mathcal{O}_i-\mathcal{O}_0\right)$ are $\pm(n+1)$ and the corresponding eigenstates are $\ket{\Psi_n}$ and $\ket{\Psi^{\perp}_n}$ respectively. Therefore, only for $\ket{\Psi_n}$ and $\ket{\Psi^{\perp}_n}$, the maximum algebraic value of $\mathcal{O}$ {\em i.e.}, $\mathcal{O}$ is equal to $n+1$, and hence violets the inequality (\ref{GHZPnI}) maximally. Thus, for odd $n$ the relations given in (\ref{GHZPn}) uniquely determines the correlation of $\ket{\Psi_n}$. Therefore, if the product of the local measurement results associated to the observable $\mathcal{O}_{j_r}$ is equal to the eigenvalue $\lambda_{j_r}$ then the correlation is a genuine n-qubit GHZ correlation.
\section{Conclusion} In conclusion, we present secure quantum protocols for both the Dining Cryptographers (DC) problem and the Anonymous Veto (AV) problem. The security of these protocols are based on GHZ paradox and the properties of the GHZ correlation. In our DC protocol, multiple payments can be detected whereas no classical protocol has this luxury with an unconditional security proof. We then generalize the DC problem for $n$ members based on the assumption that no multiple payments would occur. By exploiting this generalized DC problem we have shown that the multi-qubit GHZ state allow us to find a simple solution for the Anonymous Veto problem.
\section{Acknowledgments} We thank Sibasish Ghosh, Marek \.{Z}ukowski and Marcin Wie\'{s}niak for various discussions and comments. R.R. acknowledges support from UGC (University Grants Commission, Govt. of India) sponsored Start-Up Grant.

\end{document}